\documentclass[a4paper,11pt]{article}
\pdfoutput=1
\usepackage[T1]{fontenc}
\usepackage[utf8]{inputenc}
\usepackage{jheppub}
\usepackage{slashed}

\title{Identifying sea and valence quarks in a magnetically driven catalysis}

\author{Daniel Kosoi}
\author{and Leonardo Pati\~no}

\affiliation{Departamento de F\'{\i}sica, Facultad de Ciencias,
Universidad Nacional Aut\'onoma de M\'exico,\\
Apartado Postal 70-542, CP 04510, Ciudad de M\'exico, M\'exico}

\emailAdd{danielkosoi@ciencias.unam.mx}

\abstract{In the present work, we study the magnetic response of a holographic plasma in a D3/D7 setup through the quasinormal spectrum of D7-brane embedding fluctuations. Extending the spectral analysis to nonzero spatial momentum, we show that the modes become increasingly quasiparticle-like as the kinetic momentum grows at fixed magnetic field, while increasing the field at fixed kinetic momentum suppresses this character. The possible existence of these competing effects in an actual quark-gluon plasma could account, on the one hand, for the partial success of quasiparticle-based effective descriptions due to the presence of a substantial population of modes stabilized by their high momentum, and, on the other, for the difficulty of isolating experimental signatures of such constituents because of the magnetic suppression that dominates at low momentum. Within the same setup, we introduce a holographic prescription to describe the contribution of valence and sea quarks to the phenomena of magnetic catalysis and its inverse, characterized herein through the induced expectation value of a scalar operator. We apply this prescription at vanishing spatial momentum to perform an initial calculation leading to results that are in agreement with those obtained from lattice QCD.}

\keywords{Gauge-Gravity Correspondence, D-Branes, Quark-Gluon Plasma}

\begin{document}
\maketitle
\flushbottom

\section{Introduction and main results}
\label{sec:intro}

Since its first form, speculated in the 1970s, the phase diagram of Quantum Chromodynamics (QCD) has become increasingly more complex, considering dependencies on parameters other than temperature and baryon chemical potential, such as the masses of the quarks or separated the chemical potential for each flavor. These and other elements have been incorporated \cite{Karch:2002sh,Kobayashi:2006sb,Erdmenger:2008yj} into the description that is conjectured to be provided by the gauge/gravity correspondence \cite{Maldacena:1997re}. The motivation to use such tool is that there are regions of the aforementioned phase diagram that escape the usual perturbative treatment of quantum field theory, making the so called holographic calculations an appealing alternative. Another agent that has proven to have a relevant influence on the shape of the phase diagram is an intense external magnetic field, expected to be generated in non-central high energy collision experiments where the state of matter known as quark-gluon plasma (QGP) is produced. Even the most conservative estimations predict an intensity of $10^{-1}{m_{\pi}}^2$ for this magnetic field, while the more extreme go as high as $15{m_{\pi}}^2$ \cite{Skokov:2009qp}. Regardless of the specific value, any intensity in this range is guaranteed to have consequences not only on the phase diagram, but also importantly, on the interpretation of experimental results, so a lot of effort has been placed to understand the related phenomenology.

A five-dimensional background was constructed in \cite{DHoker:2009mmn} to model a strongly coupled plasma subjected to an external magnetic field, and was later uplifted to ten dimensions in \cite{Elinos:2021bmx} to incorporate fields in the fundamental representation. Back then we showed that the quasinormal modes of a fundamental scalar operator organized themselves into Landau-like levels at vanishing spatial momentum. In the present letter, we extend our study in \cite{Elinos:2021bmx} to spectral functions with non-zero spatial momentum, allowing us to extract the dispersion relation of the modes and therefore analyze how closely they resemble quasiparticles. Our construction permits us to separate the effect of kinetic momentum from the one of the magnetic field, showing that increasing the former enhances the quasiparticle character of the modes, whereas increasing the latter suppresses it. If these competing effects were present in the real world quark-gluon plasma, it would be appealing to think that the scenarios where effective quasiparticle models have been successful are those involving the substantial population of modes that at high temperatures are in the quasiparticle regime due to their high momentum, while the magnetic suppression dominating at low momentum would explain the difficulty to isolate experimental signatures of quasiparticle constituents.

Turning back to spectral functions with zero spatial momentum, we use the same holographic setup to probe separately the effect of the magnetic field on the backreacted background geometry and its direct effect on the perturbation. We pursue this separation to study what we, based on the field theory expressions in \cite{Bruckmann:2013}, identify in our model as the gravitational duals of the sea and valence contributions to the response of the chiral condensate to a strong magnetic field, and to test this proposal by determining if such contributions develop the distinctive behaviors reported in the aforementioned work.

In our prescription the sea contribution is encoded in the backreaction of the background geometry to the magnetic field, while the valence one lies in the direct dependence on said field of the flavor-brane perturbations. From this identification we notice that, even though dynamical sea-quark contributions are absent due to the quenched approximation \cite{Bruckmann:2013} inherent here because of the probe limit of the embedded objects, the way the magnetic field is incorporated into the gravitational model allows us to report some static effects of the sea-quarks, since it affects the metric induced on the D7-branes.

We show that the valence response depends on the canonical momentum and magnetic field only through the way they combine in the kinetic momentum, and then proceed to use the induced scalar expectation value to characterize the magnetic response, identifying regimes of enhancement and suppression across different ranges of field strength, analogous to magnetic catalysis and inverse magnetic catalysis of the chiral condensate. In particular, we find that the valence contribution consistently enhances the induced expectation value, while all suppression originates from the sea contribution.

To complement this analysis, in the appendix we examine the widths and characteristic energies of the quasinormal modes as functions of the magnetic field. We find that increasing the field tends to enhance their quasiparticle character through the valence contribution and suppress it through the sea contribution, with the full response reflecting the competition between these effects.

\section{Gravitational model and previous results}
\label{sec:Uplift}  
To provide a gravitational dual of a strongly coupled plasma with degrees of freedom in the fundamental representation subject to a constant magnetic field, in \cite{Elinos:2021bmx} we constructed the ten dimensional uplift of the five dimensional background introduced in \cite{DHoker:2009mmn}, and embedded a D7-brane on it. This was done in such a manner that the fundamental degrees of freedom were massless, and left us proceeded to study the perturbations of the brane that are dual to scalar excitations of said fundamental fields in the gauge theory.

Our results in \cite{Elinos:2021bmx} were obtained using either gauge, Landau $\mathbf{A}=B\, x\, dy$ or symmetric $\mathbf{A}=B/2(\, x\, dy-y\, dx)$, to introduce the magnetic field $\mathbf{B}=d\mathbf{A}$, but for the sake of concreteness, we will employ the former in what follows. Once Landau gauge has been adopted, the ten dimensional metric, consistent with the symmetries introduced in  \cite{DHoker:2009mmn}, reads
{\small
\begin{equation}
\begin{split}
&ds_{10}^2=-U(r)dt^{2} + \frac{1}{U(r)}dr^{2}+V(r) \left( dx^{2} + dy^{2} \right)+ W(r)dz^{2} \\
&+ \left[ d\theta ^{2}+\sin^{2}\theta d\widetilde{\phi}_1^2 + \cos^{2}\theta\left( d\vartheta^{2} +\sin^{2}\vartheta d\widetilde{\phi}_2^2+\cos^2\vartheta d\widetilde{\phi}_3^2\right)  \right],    
\end{split}\label{ds10ds5}
\end{equation}}
with $d\widetilde{\phi}_i=d\phi_i+\frac{2}{\sqrt{3}}B\, x\, dy$. The directions $t, \ x, \ y$ and $z$ are dual to those of the spacetime where the field theory lies, while $r$ is the radial holographic coordinate, on which the functions $U(r), \, V(r),$ and $W(r)$, depend solely, and the compact directions are described by the second line in eq. \eqref{ds10ds5}.

For stability reasons \cite{Elinos:2021bmx}, the D7-brane is embedded to extend in the $\{r, t, x, y,z \}$ directions and wrap the 3-cycle at the end of eq. \eqref{ds10ds5} given by $\vartheta, \phi_2,$ and $\phi_3$. In this manner, the embedding can be described by placing the brane at constant $\phi_1$ and writing $\theta$ as a function of $r$. From eq.  \eqref{ds10ds5} we notice that, as usual in these descriptions, it is convenient to use $\chi(r)=\sin[\theta(r)]$ to study the profile of the brane. The object of interest to our work are the perturbations $\delta\chi$ to the aforementioned embedding, of which in \cite{Elinos:2021bmx} we took the particular case $\chi(r)=0$ so that the resulting equation for the brane perturbations, when expressed as the product $\delta\chi(r,t,x,y,z)=\chi_t(t)\chi_x(x)\chi_y(y)\chi_z(z)\chi_r(r)/\chi_r(r_\infty)$, reduces to
\begin{equation}
\begin{split}
& \bigg[3 U V W' \chi_r'+6 W \big(V U' \chi_r'+U V' \chi_r'+U V \chi_r''\big)+6 V W\left(3 -\frac{\partial_t^2 \chi_t}{U \chi_t}+\frac{\partial_y^2 \chi_y}{V \chi_y}+\frac{\partial_z^2 \chi_z}{W \chi_z} \right) \chi_r\bigg]\chi_x\\
&+W \chi_r \left(6 \partial_x^2\chi_x-8 B^2 x^2 \chi_x\right)=0,
\end{split}
\label{eompsirx}
\end{equation}
with a general solution that includes the factor $\chi_t(t)\chi_y(y)\chi_z(z)=e^{-i (\omega\, t-k_y y-k_z z)},$ and which can be further separated as
\begin{equation}
\begin{split}
&\left(3 +\frac{\omega^2}{U}-\frac{k_y^2}{V}-\frac{k_z^2}{W}  \right) V W \chi_r+\frac{1}{2} U V W' \chi_r'+ W \left(V U' \chi_r'+U V' \chi_r'+U V \chi_r''\right)= 2 {\cal{E} } W \chi_r,\end{split}
\label{eompsirL}
\end{equation}
and
\begin{equation}
\frac{1}{2}\left[-\partial_x^2\chi_x+e^2 B^2 x^2 \chi_x\right]={\cal{E}} \chi_x .\label{eompsix}
\end{equation}
In the expressions above $e=\frac{2}{\sqrt{3}}$, and we have taken into account that $U(r), \, V(r),$ and $W(r)$, are only functions of the radial coordinate $r$, denoting the differentiation with respect to the latter by a prime.

The details of the ten dimensional background, and the embedding of the brane, can be found in \cite{Elinos:2021bmx}, but for completeness we should mention that there is a 5-form that plays a relevant role in the construction of the uplift, but has no effect on the embedding nor on our current calculations, and that the 1-forms $d\widetilde{\phi}_i=d\phi_i+\frac{2}{\sqrt{3}}B\, x\, dy$ in eq.  \eqref{ds10ds5} show that, as the $U(1)$ field is encoded in these compact directions, the quantity $e=\frac{2}{\sqrt{3}}$ correctly represents the charge with respect to such fields.

We began our previous study of the equations above noticing that for the solutions to describe acceptable embeddings, the separation constant ${\cal{E}}$ had to take values on the discrete spectrum
\begin{equation}
{\cal{E}}_n=\left( n+\frac{1}{2}\right)\omega_c, \label{LL1}
\end{equation}
where $\omega_c=e\,B$ was identified as the cyclotron frequency and the integer $n$ with the energy level number. 

We continued by focusing on the $k_y=k_z=0$ case, and looked for the complex values of $\omega$ for which the solutions to eq.  \eqref{eompsirL} were ingoing and had a normalizable profile in the radial direction, since these are dual to quasinormal modes of the scalar excitations \cite{Kovtun:2005ev}.

It was the energy obtained from the latter frequencies that we proved to closely follow the dispersion relation characteristic of Landau levels, presenting a small deviation as expected from modes that are not fully stable. Furthermore, when we studied the ratio of the width of the states over their energy as a function of the magnetic field, we observed a behavior reminiscent of magnetic catalysis for large intensities of such field, and inverse magnetic catalysis for small ones, all with respect to an identifiable critical intensity.

Isolating the effect of the magnetic field was not possible in this previous work, where the calculations were made at vanishing spatial momentum and as the intensity of said field became larger, its direct impact was certainly augmented, but also unavoidably increased the kinetic momentum associated with each level.

While the induced scalar expectation value is studied here at $k_z=0$, we extend the spectral analysis to nonzero spatial momentum in eq. \eqref{eompsirL} to distinguish the effects of kinetic momentum and magnetic field on the quasiparticle properties of the modes. For the $n$-th level, the kinetic momentum satisfies ${k_K}^2=2e\, b\,(n+1/2)+k_z^2$. At fixed $n$, allowing $k_z$ to vary therefore enables us to study the quasinormal modes as functions of the magnetic field strength over the range $0\leq b\leq {k_K}^2/ 2e\, (n+1/2)$, while keeping the kinetic momentum constant by decreasing $k_z$ from $k_K$ to zero as $b$ increases.

\section{Finding the quasinormal frequencies}

As can be consulted in \cite{Elinos:2021bmx}, close to the horizon the metric functions are given by the expansions
\begin{equation}
\begin{split}
&U(r)=6 r_h (r-r_h)+\sum _{\text{i}=2}^\infty U_i(r-r_h)^i,\\
&V(r)=\sum _{\text{i}=0}^\infty V_i(r-r_h)^i,\\
&W(r)=3\,{r_h}^2 (r-r_h)^0 +\sum _{\text{i}=1}^\infty W_i(r-r_h)^i,
\end{split}\label{seriesUVW}
\end{equation}
which through eq. \eqref{eompsirL} show that the radial profile of the embedding behaves like $(r-r_h)^{i\alpha}$, with $\alpha=\pm \frac{\omega}{4\pi T}=\pm \frac{\omega}{6 r_h}$, in the $r\rightarrow r_h$ limit. The ingoing wave requirement is imposed by choosing the negative sign for $\alpha$ and approximating $\chi^{(\omega,n)}_r(r)$ near the horizon through the resulting series
\begin{equation}
\begin{split}
&\chi^{(\omega,n)}_r(r)\simeq(r-r_h)^{-i\frac{\omega}{6 r_h}} \chi_r^{(0)}\left[1+C_{(r,\omega,n)}^{(1)}(r-r_h)+C_{(r,\omega,n)}^{(2)}(r-r_h)^2+{\mathcal{O}}(r-r_h)^3\right],
\end{split}
\label{seriespsi}
\end{equation}
where $\chi_r^{(0)}$ is a free global factor due to the linear character of eq. \eqref{eompsirL}, while
\begin{equation}
\begin{split}
    &C_{(r,\omega,n)}^{(1)}=\frac{1}{108 {r_h}^2 V(r_h)^2 (3 {r_h} - i\omega)}\big\{b^2\omega (5\omega - 3 i r_h)+18k_z^2V(r_h)^2\\ &- 6 V(r_h)\left[-18 (n+1/2)e\, b {r_h}^2  + V(r_h)\left (27 {r_h}^2 - 15 i {r_h}\omega + \omega^2 \right) \right]\big\}\\
    &C_{(r,\omega,n)}^{(2)}=\frac{1}{23328 {r_h}^4 V(r_h)^4 \left(18
   {r_h}^2-9 i {r_h} \omega -\omega ^2\right)}\\
   &\Big\{b^4 \omega  \left(-252 {r_h}^2
   \omega -468 i {r_h}^3+35 i {r_h} \omega ^2+25 \omega ^3\right)+324 k_z^4V(r_h)^4\\
   &+ 12 b^2 V(r_h) \Big[18 (n+1/2)e\, b {r_h}^2 
   \left(36 {r_h}^2-12 i {r_h} \omega +5 \omega ^2\right)\\
   &+V(r_h) \left(-459 {r_h}^2 \omega ^2-288 i {r_h}^3
   \omega +972 {r_h}^4+97 i {r_h} \omega ^3-5 \omega ^4\right)\Big]+\\
   &36 V(r_h)^2 \Big[324 (n+1/2)^2 e^2\, b^2 {r_h}^4 -36 (n+1/2)e\, b {r_h}^2 V(r_h)\\
   &\left(99  {r_h}^2-24 i {r_h} \omega +\omega ^2\right)\\
   &+V(r_h)^2 \left(-126 {r_h}^2 \omega ^2-2268 i {r_h}^3 \omega
   +2673 {r_h}^4-45 i {r_h} \omega ^3+\omega ^4\right)\Big]\\
   &-36k_z^2V(r_h)^2[B^2(72r_h^2-24ir_h\omega-5\omega^2)\\
   &+6V(r_h)(-18(n+\frac{1}{2})e b r_h^2+V(r_h)(99r_h^2-24ir_h\omega+\omega^2))]\Big\}.
\end{split}
\label{coefseriespsi}
\end{equation}

In the asymptotic region $r\rightarrow\infty$ the metric functions are described by
\begin{equation}
\begin{split}
    &U(r)=r^{2}+U_{1}r+\frac{U_{1}^{2}}{4}+\frac{1}{r^{2}}\left(U_{-2}-\frac{2}{3}b^{2}\log{r}\right)+U_1\frac{1}{r^{3}}\left(-U_{-2}-\frac{1}{3}b^2+\frac{2}{3}b^{2}\log{r}\right)+\mathcal{O}\left(\frac{1}{r^{4}}\right),\\
    &V(r)=r^{2}+U_{1}r+\frac{U_{1}^{2}}{4}+\frac{1}{r^{2}}\left(V_{-2}+\frac{1}{3}b^{2}\log{r}\right)+U_1\frac{1}{r^{3}}\left(-V_{-2}+\frac{1}{6}b^2-\frac{1}{3}b^{2}\log{r}\right)+\mathcal{O}\left(\frac{1}{r^{4}}\right),\\
    &W(r)=r^{2}+U_{1}r+\frac{U_{1}^{2}}{4}+\frac{1}{r^{2}}\left(-2V_{-2}-\frac{2}{3}b^{2}\log{r}\right)+U_1\frac{1}{r^{3}}\left(2V_{-2}-\frac{1}{3}b^2+\frac{2}{3}b^{2}\log{r}\right)+\mathcal{O}\left(\frac{1}{r^{4}}\right),
\end{split}\label{BseriesUVW}
\end{equation}
and the radial profile therefore by
\begin{equation}
\begin{split}
&\chi^{(\omega,n)}_r(r)\simeq {\chi_r}^{(-1)}\bigg[\frac{1}{r}-\frac{U_1}{2r^2}+\left(\omega^2-k_z^2-2 (n+1/2)e\, b\right)\frac{\log{r}}{2 r^3}-3 U_1 \left(\omega^2-k_z^2-2 (n+1/2)e\, b\right)\frac{\log{r}}{4 r^4}\\
&+ U_1 \left(\omega^2-k_z^2-2 (n+1/2)e\, b+{U_1}^2\right)\frac{1}{4 r^4}\bigg]+{\chi_r}^{(-3)}\left[\frac{1}{r^3}-6 U_1 \frac{1}{4 r^4}\right]+\mathcal{O}\left(\frac{1}{r^{5}}\right),
\end{split}
\label{Bseriespsi}
\end{equation}
as can be verified by substituting eq. \eqref{BseriesUVW} into eq.  \eqref{eompsirL}.

From the holographic dictionary, presented in \cite{Myers:2007we}, we see that the expansion coefficient $\chi_r^{(-1)}$ is related to the source of the scalar operator in the dual field theory as 
\begin{equation}
\mu(\omega,b)=\frac{1}{2}\sqrt{\lambda}\chi_r^{(-1)}(\omega,b),
\end{equation}
while the coefficient, $\chi_r^{(-3)}$, is proportional to the induced expectation value \cite{Myers:2007we} of the scalar operator in a manner which will be described in a following section. The quasinormal modes associated with the excitations dual to the brane embedding fluctuations are given by normalizable solutions, identified as those for which $\chi_r^{(-1)}$ vanishes, since, as seen in eq. \eqref{Bseriespsi}, this is the coefficient that multiplies the non-normalizable part of the radial profile. 

The first task is to obtain the quasinormal frequencies as functions of the dimensionless parameters $b/T^2$ and ${k_K}^2/T^2$. To this end,  we notice that the specific behavior of $\chi(r)$ over any member of our family of backgrounds, at a given temperature $T$ and intensity $b$ of the magnetic field, is parametrized by $\omega$, $k_y$, $k_z$ and $n$. We fix our attention on the lowest energy level set by $n=0$ in eq.  \eqref{LL1}, and choose $k_y=0$ in eq. \eqref{eompsirL} while employing $k_z$ as described above to select a value for ${k_K}^2=e\, b+{k_z}^2$. Numerically solving eq. \eqref{eompsirL} under these circumstances and near horizon conditions given by eq. \eqref{seriespsi} for a particular value of $\omega$, permits us to use the asymptotic behavior of the solution $\chi(r)$ to extract the coefficients $\chi_r^{(-1)}(\omega)$ and $\chi_r^{(-3)}(\omega)$ corresponding to this frequency.  Performing such integration for values of $\omega$ that explore the complex plane, starting at the origin and searching for the nearest locus $\operatorname{Re}[\chi_r^{(-1)}(\omega)]=0$ and $\operatorname{Im}[\chi_r^{(-1)}(\omega)]=0$, leads us to extract the frequency of the quasinormal mode as the value at which these lines intersect. Repeating the procedure above for several $k_K$'s over a series of backgrounds with a range of temperatures and intensities for the magnetic field renders the desired function $\omega(b/T^2,{k_K}^2/T^2)$.

For the calculation of the induced scalar expectation value in sections \ref{CCond} and \ref{sec:IndExpResults}, we follow the described procedure at $k_z=0$, and extract the asymptotic value at the quasinormal frequency $\omega_0$ as $\chi^{(-3)}(b/T^2)$. From the linearity of eq. \eqref{seriespsi}, it follows that the radial solution is defined up to a multiplicative constant. This normalization freedom is inherited by both asymptotic expansion coefficients, and to extract physically meaningful results, the usual practice is to fix one of them to a finite constant while studying the behavior of the other. The latter procedure is not applicable in our case, since we precisely determine the quasinormal frequency $\omega_0$ as the one for which the source term vanishes. Nonetheless, close to such value of the frequency we can write
\begin{equation}
\chi_r^{(-1)}(\omega)=(\omega-\omega_0)\partial_{\omega}\chi_r^{(-1)}(\omega)|_{\omega=\omega_0}+\mathcal{O}(\omega-\omega_0)^2,
\end{equation}
where the constant coefficient $\partial_{\omega}\chi_r^{(-1)}(\omega)|_{\omega=\omega_0}$ is closely related to the residue of the quasinormal mode \cite{Kaminski:2009dh} and proportional to the same aforementioned overall factor. We therefore see that a direct comparison of the values of $\chi_r^{(-3)}$ remains meaningful among solutions with vanishing source terms, just as long as all of them have the same value for $\partial_{\omega}\chi_r^{(-1)}(\omega)|_{\omega=\omega_0}$. In what follows we will set this constant to 1, which in practice, given the linearity of \eqref{seriespsi}, amounts to finding any solution and dividing it by the value that $\partial_{\omega}\chi_r^{(-1)}(\omega)|_{\omega=\omega_0}$ acquires in it.

\section{Relation to the quasiparticle models}
\label{QPModel}

A quasiparticle model of the quark-gluon plasma \cite{Peshier:1996} has been successfully used to describe certain aspects of it \cite{Cassing:2008}, including scenarios where it is magnetized \cite{Kurian:2019}.

Before we study the behavior of the condensate, it is interesting to see that the quasinormal modes in our work follow a dispersion relation consistent with said model. To this end, we would like to display the energy of the mode as a function of the kinetic momentum at constant magnetic field. Since the kinetic momentum $k_K$ is defined by the relation ${k_K}^2=eb+{k_z}^2$, isolating the effect that it has on the quasinormal modes is a simple task, as it suffices to explore the result of changing $k_z$ while keeping $b$ constant. To graphically compare the results for different values of $b/T^2$, it turns out to be convenient to compensate for the shift that the field introduces, so the results will be plotted as functions of $({k_K}^2-eb)/T^2$.

\begin{figure}[htbp]
    \centering
    \includegraphics[width=0.75\textwidth]{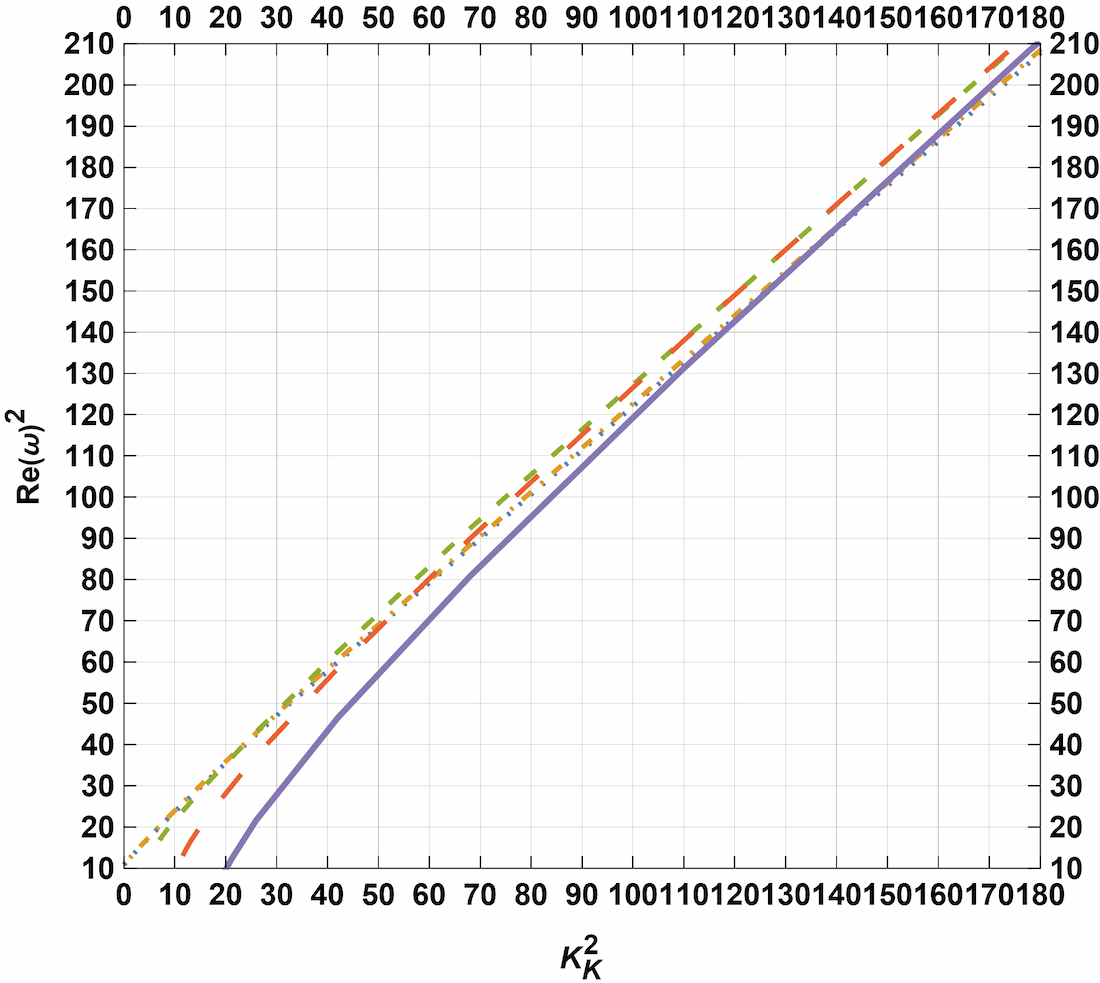}
    \caption{Dispersion relation associated with the quasinormal states. The dotted, dot dashed, dashed (medium),  dashed (large) and continuous curves correspond to $b/T^2=\{ 0, 4{\pi}^2/9, 8{\pi}^2/3, 40{\pi}^2/9, 64{\pi}^2/9\} $ respectively.}
    \label{DispRel}
\end{figure}

\begin{figure}[htbp]
    \centering
     \includegraphics[width=0.75\textwidth]{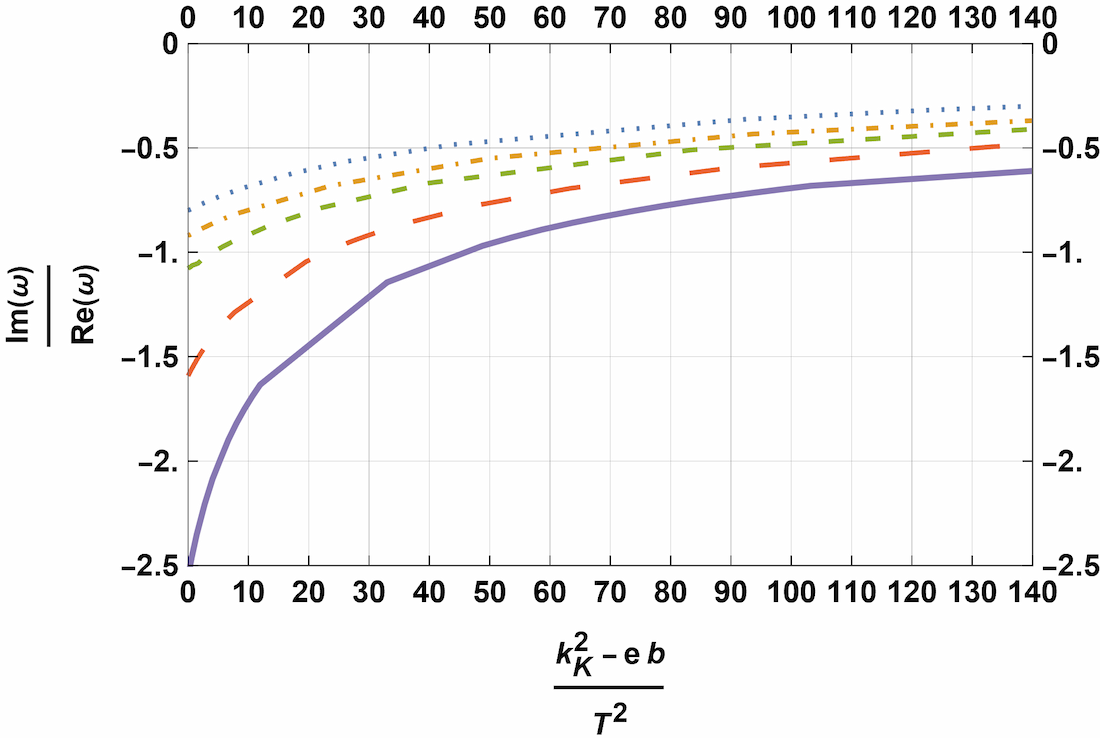}
    \caption{Ratio of the imaginary to real parts of the quasinormal frequency as a function of the kinetic momentum
    $({k_K}^2-eb)/T^2$ at fixed magnetic field intensity. The dotted, dot dashed, dashed (medium), dashed (large) and continuous curves correspond to $b/T^2=\{ 0, 4{\pi}^2/9, 16{\pi}^2/3, 40{\pi}^2/9, 64{\pi}^2/9\} $ respectively.}
    \label{IcR-b}
\end{figure}

In figure \ref{DispRel} we show the energy of the mode, given by the real part of the quasinormal frequency, as a function of the kinetic momentum. We notice that it is monotonically increasing and becomes linear for large values of ${k_K}^2/T^2$. One can observe that the dispersion relation for a massless transverse field in a thermal theory \cite {Bellac:2011kqa}, characterized by the asymptotic behaviors \begin{equation}
\begin{split}
\text{Re}[\omega]^2\simeq {\omega_p}^2+\alpha {k_K}^2\quad \text{for} \ \ k_K\ll eT,\\
\mathrm{and}\\
\text{Re}[\omega]^2\simeq {m_T}^2+ {k_K}^2\quad \text{for} \ \ k_K\gg eT,
\end{split}\label{Asym}
\end{equation}
is recovered, providing a strong indication that the quasinormal modes are indeed acting like quasiparticles in the plasma. To confirm this observation, in figure \ref{IcR-b} we compare the real and imaginary parts of their frequency and notice that, as the kinetic momentum grows, the former surpasses the latter in magnitude, making the quasiparticle comparison adequate \cite{Myers:2007we,Hoyos-Badajoz:2006dzi}.

\begin{figure}[htbp]
    \centering
    \includegraphics[width=0.75\textwidth]{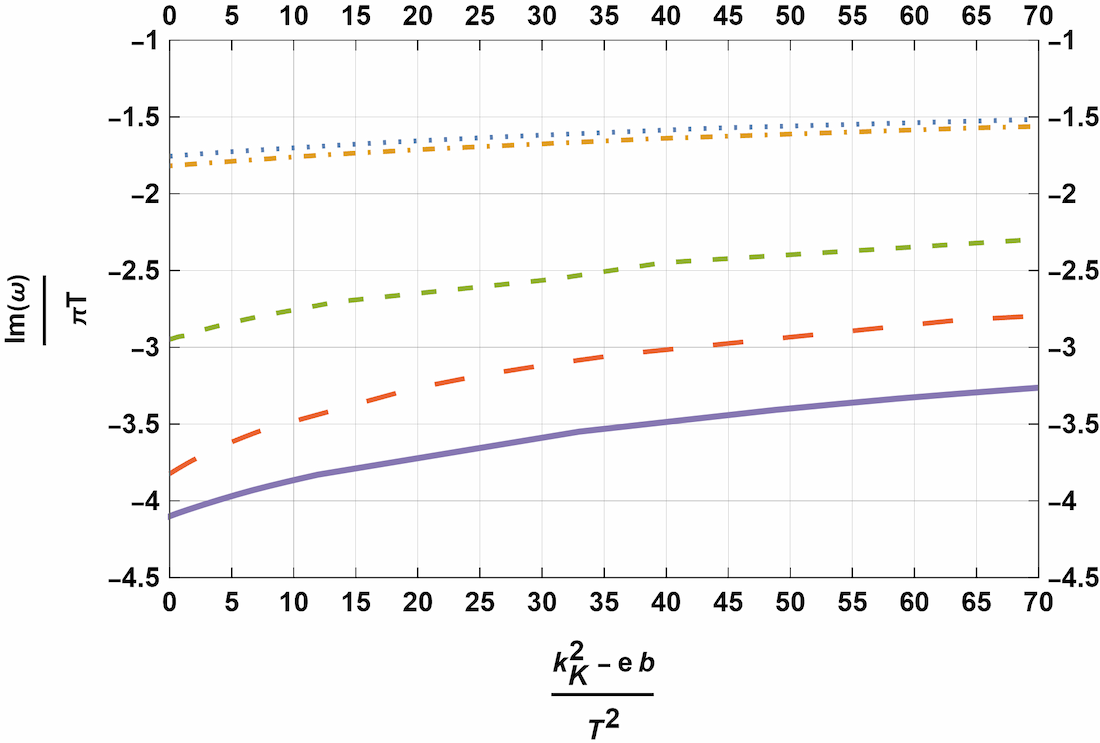}
    \caption{Imaginary part of the quasinormal frequency as a function of the kinetic momentum
    $({k_K}^2-eb)/T^2$ at fixed magnetic field intensity. The dotted, dot dashed, dashed (medium),  dashed (large) and continuous curves correspond to $b/T^2=\{ 0, 4{\pi}^2/9, 8{\pi}^2/3, 40{\pi}^2/9, 64{\pi}^2/9\} $ respectively.}
    \label{Im-K}
\end{figure}

Additionally, figure \ref{Im-K} shows that the imaginary part of the quasinormal frequency decreases in magnitude as a function of the kinetic momentum, indicating that this quantity has a stabilizing effect on the quasiparticle-like states.

As mentioned before, for any given value $k_K$ of the kinetic momentum and not to exceed it, the magnetic field intensity can only be in the interval $0\leq b\leq {k_K}^2/e$. To be able to make comparisons for plots at different values of fixed kinetic momentum, we normalize the horizontal axis by the maximum value that $b$ can take. 

\begin{figure}[htbp]
    \centering
    \includegraphics[width=0.75\textwidth]{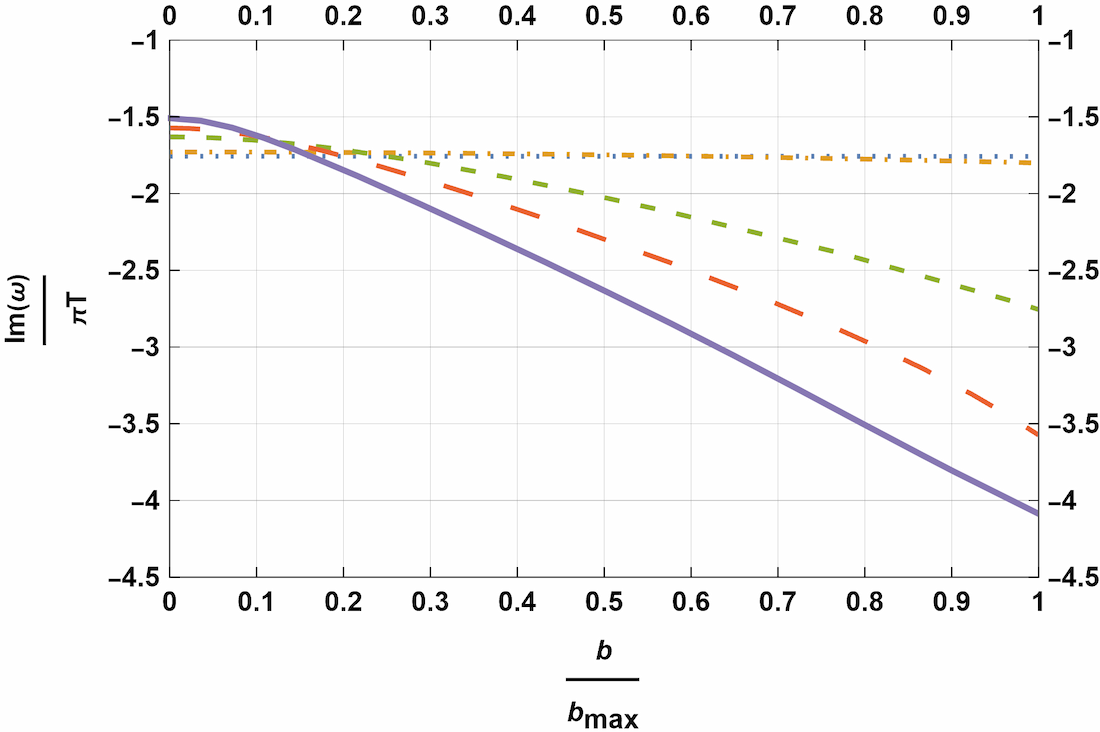}
    \caption{Imaginary part of the quasinormal frequency as a function of the magnetic field $b/T^2$ at fixed kinetic momentum. The dotted, dot dashed, dashed (medium), dashed (large) and continuous curves correspond to ${k_K}^2/T^2=\{ 2{\pi}^2/45, 4{\pi}^2/9, 8{\pi}^2/3, 40{\pi}^2/9, 64{\pi}^2/9\} $  respectively.}
    \label{IcV-b}
\end{figure}

In figure \ref{IcV-b} we show the dependence of the width $\operatorname{Im}[\omega]$ on $b/b_{max}$ at fixed ${k_K}^2/T^2$. The curves are monotonically increasing for every value of the momentum, reinforcing that the magnetic field has a destabilizing effect on the quasiparticle-like states. It is of note that as ${k_K}^2$ grows, the graphs accommodate from top to bottom, consistent with the notion that the kinetic momentum has a stabilizing effect. 

\begin{figure}[htbp]    
    \centering
    \includegraphics[width=0.75\textwidth]{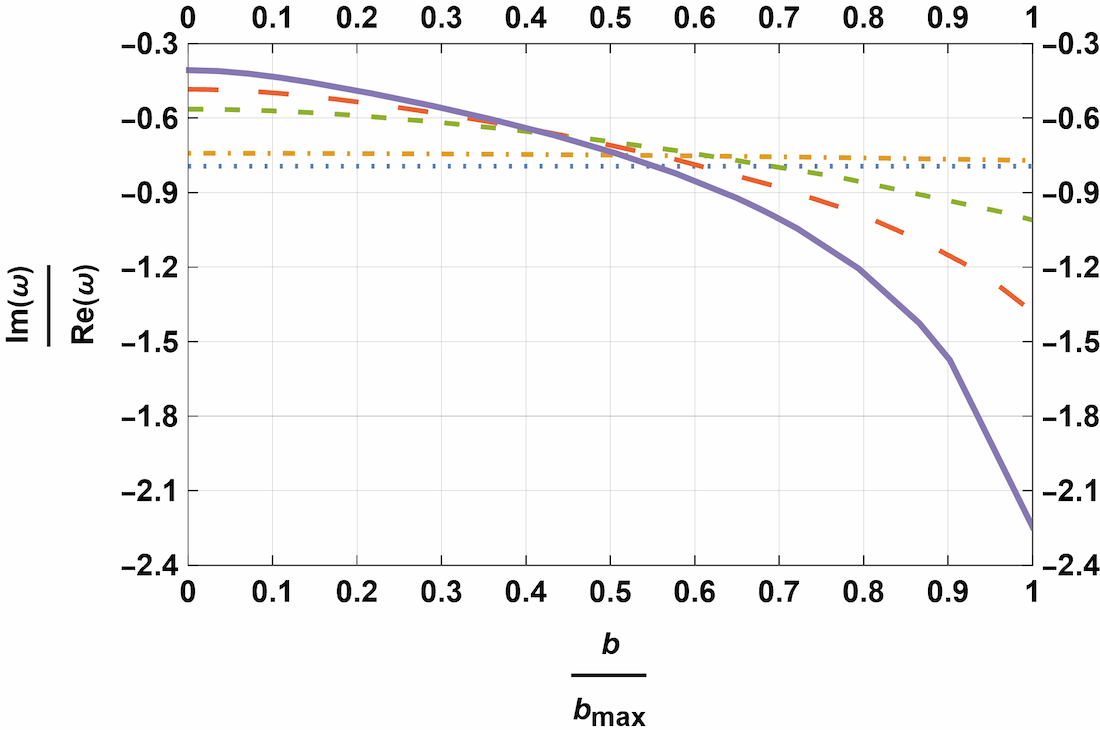}
    \caption{Ratio of the imaginary to real parts of the quasinormal frequency as a function of the magnetic field $b/T^2$ at fixed kinetic momentum. The dotted, dot dashed, dashed (medium), dashed (large) and continuous curves correspond to ${k_K}^2/T^2=\{ 2{\pi}^2/45, 4{\pi}^2/9, 8{\pi}^2/3, 40{\pi}^2/9, 64{\pi}^2/9\} $  respectively.}
    \label{IcR-K}
\end{figure}

In figure \ref{IcR-K}, we present the ratio of the imaginary to real parts of the quasinormal frequency, this time as a function of the magnetic field. The curves decrease with increasing field intensity, rendering the quasiparticle interpretation unviable under these conditions.

Part of our motivation to work at fixed kinetic momentum becomes apparent in light of the conclusions above, since had we increased $b$ without using $k_z$ to compensate the change in $k_K$, we would have seen the result of competing destabilizing and stabilizing effects, given that $b$ also augments the kinetic momentum.

Our plots suggest that, for the explored strengths of magnetic field, sufficiently large kinetic momentum can counteract magnetic suppression and restore quasiparticle-like behavior.

\section{Role of the sea and valence quarks in (inverse) magnetic catalysis}
\label{sec:ValenceSea}

The difference between sea and valence quarks is technically artificial, but has proven to be a very useful tool in investigations concerning the origin of (inverse) magnetic catalysis. It is worth noting that this phenomenon can refer either to the decrease of the chiral critical temperature \cite{Preis:2010cq,Ayala:2014yla} or to the suppression of the chiral condensate \cite{Bali:2012zg,Bruckmann:2013} as a function of the magnetic field intensity; our focus will be on the latter. The aim of the present letter is to exploit this distinction to identify the elements of our gravitational construction that encode the two distinct effects of the magnetic field on the induced scalar expectation value. 

In \cite{Bruckmann:2013} a way to separate and explore each effect by itself is presented. Unlike our case, in said work the field under consideration is fermionic, and its condensate is determined to be given by
\begin{equation}
\begin{split}
&\overline{\psi}\psi(b)=\frac{1}{\mathcal{Z}(b)}\int\mathcal{D}Ue^{-S_g}det(\slashed{D}(b)+m)\mathrm{Tr}(\slashed{D}(b)+m)^{-1},\\
&\mathrm{where}\\
&\mathcal{Z}(b)=\int\mathcal{D}Ue^{-S_g}det(\slashed{D}(b)+m).
\end{split}
\label{CT}
\end{equation}

From this expression, the valence and sea contributions to the condensate are defined as
\begin{equation}
\begin{split}
    &\overline{\psi}\psi^{val}(b)=\frac{1}{\mathcal{Z}(0)}\int\mathcal{D}Ue^{-S_g}det(\slashed{D}(0)+m)\mathrm{Tr}(\slashed{D}(b)+m)^{-1},\\
    &\overline{\psi}\psi^{sea}(b)=\frac{1}{\mathcal{Z}(b)}\int\mathcal{D}Ue^{-S_g}det(\slashed{D}(b)+m)\mathrm{Tr}(\slashed{D}(0)+m)^{-1},
\end{split}
\label{CVS}
\end{equation}
reflecting that, in the presence of a magnetic field, the valence effect is encoded in a shift of the low-lying modes of the Dirac operator—given by the Banks–Casher relation \cite{Banks:1979yr}—while the sea effect modifies the condensate through the fermion determinant.

It has also been shown, using lattice QCD techniques, that the valence contribution always enhances the condensate, leading to magnetic catalysis \cite{DElia:2010abb,DElia:2011koc}, and constitutes a universal feature of the system \cite{Endrodi:2013cs}. In contrast, the sea effect is responsible for the suppression of the condensate, giving rise to inverse magnetic catalysis, and is entirely absent in the quenched approximation. This contribution becomes significant only near the critical temperature and is independent of the central value around which the gauge field is sampled.

From the holographic perspective, the fact that we are working in the probe approximation implies that the dual flavors are quenched, so the corresponding fermion determinant is unaffected by the fields on the D7-brane. Nonetheless, the magnetic field in our setup is not part of the worldvolume content but of the background itself, so the geometry fully backreacts to it, and along with it, the embedding and induced metric on the D7-brane. This means that even if the dynamical processes of the sea quarks are not captured by our construction because they are quenched, we can still explore the static effect that changing the magnetic field has on the sea contribution to the induced expectation value, since such change will modify the determinant.

Concretely, changing $b$ has a direct and explicit impact on eq. \eqref{eompsirL}, but it also modifies the equation indirectly through the metric functions $U(r)$, $V(r)$, and $W(r)$, whose radial profiles depend on the intensity of the magnetic field and consequently determine the embedding and induced metric of the D7-brane.

Motivated by the above and by the distinction introduced in \cite{Bruckmann:2013}, we will employ our construction to identify the shift in the spectrum as dual to the effect that the explicit appearance of $b$ in eq. \eqref{eompsirL} has on the perturbation $\delta\chi$, while the change in the integration measure will be associated with the one experienced by the induced metric on the D7-branes as the bulk geometry backreacts to different intensities of the magnetic field, codified in the metric functions.

In the scalar case that we are working on, the correspondence above amounts to stating that the valence contribution to the induced expectation value ${{\langle\mathcal{O}}_{\chi}\rangle}^{\text{val}}(b)$ should be holographically computed using the coefficient $\chi^{(-3)}(b)$ as extracted from the solution to eq. \eqref{eompsirL} written with $b\neq 0$, but featuring the metric functions of a black D3-brane at $b=0$, $U_0, V_0,$ and $W_0$.

Conversely, for the sea contribution ${{\langle\mathcal{O}}_{\chi}\rangle}^{\text{sea}}(b)$ the coefficient $\chi^{(-3)}(b)$ that should be used in our holographic calculation is the one extracted from the solution to eq. \eqref{eompsirL} with $b=0$, but provided with the metric functions $U, V,$ and $W$ that have fully backreacted to the presence of the magnetic field.

The procedure described above to compute the valence expectation value simplifies eq.  \eqref{eompsirL} to
\begin{equation}
 \begin{split}
 &[3r^3(r^4-1)+r^5\omega^2]\chi_r+(1-6r^4+5r^8)\chi_r'+r(r^4-1)^2\chi_r''=r(r^4-1)(2\mathcal{E}+{k_z}^2)\chi_r,
 \end{split}
 \label{Seom}
 \end{equation}
where the metric functions of the black D3-brane have been used and $k_y$ has been set to zero as in previous sections.

From eq. \eqref{LL1} we see that the quantity $2\mathcal{E}+{k_z}^2$ in parenthesis on the right-hand side of eq. \eqref{Seom} is the kinetic momentum ${k_K}^2=eb+{k_z}^2$ for $n=0$, showing that the valence contribution depends on $k_z$ and $b$ only through $k_K$, and not on their independent values.

Given the above, and in order to compare the valence, sea, and (what we will henceforth refer to as) full contributions to the induced expectation value under the same conditions, with the latter corresponding to the case in which both effects of the magnetic field are present, we return to the study of spectral functions at zero longitudinal momentum, $k_z=0$. Since the general metric functions are known to us only numerically, we must resort to such methods to solve eq. \eqref{eompsirL} in the corresponding backgrounds.

\section{The induced scalar expectation value}
\label{CCond}

Once the quasinormal frequencies have been found, they can be used to determine the induced expectation value of the scalar operator $\langle\mathcal{O}_{\chi}\rangle$. Following the holographic dictionary, this quantity can be computed as the variation of the on-shell D7-brane Euclidean action with respect to the quark mass \cite{Myers:2007we,Avila:2020ved}
\begin{equation}
    \begin{split}
       &\langle\mathcal{O}_{\chi}\rangle=\frac{1}{\sqrt{g(0)}}\left(\frac{\delta S_E}{\delta m_q}\right)_{m_q=0}=\frac{1}{T\sqrt{g(0)}}\left(\frac{\delta \mathcal{F}}{\delta m_q}\right)_{m_q=0},
    \end{split}
    \label{qqcond}
\end{equation}
where the fact that the Euclidean action gives the leading contribution to the free energy density $\mathcal{F}=T\mathcal{S}_E$ was used to obtain the second equality.

The mass can be holographically obtained from the asymptotic distance between the color and flavor branes in the directions perpendicular to both of them. The profile of the D7-brane that determines this distance is described by the embedding function $\chi (r)$, which can be used to rewrite eq. \eqref{qqcond} in terms of a variation done with respect to this asymptotic value. After using eq. \eqref{eompsirL} and eq. \eqref{Bseriespsi} this procedure yields
\begin{equation}
    \begin{split}
        &\langle\mathcal{O}_{\chi}\rangle=\pi\sqrt{\frac{2}{\lambda}}\lim_{\epsilon\to 0}\left(\frac{1}{\epsilon^3}\frac{1}{T\sqrt{\lambda}}\frac{\delta \mathcal{F}}{\delta\chi} \right)=-\frac{1}{8}\sqrt{\lambda}N_f N_c T^3\chi^{(-3)},
    \end{split}
    \label{qqchi}
\end{equation}
where $\lambda$ is 't Hooft coupling, $N_f$ the number of D7-branes, $N_c$ the number of black D3-branes, and $T$ the temperature.

Given the form of eq. \eqref{eompsirL}, the solutions associated with quasinormal frequencies are in general complex. The reality condition for the embedding function can be met by pairing a solution with frequency $\omega$ with the solution of frequency $-\omega^*$, since the latter is the complex conjugate of the former \cite{Hoyos-Badajoz:2006dzi}. Their sum is therefore real and equal to twice the real part of either solution. In what follows, we assume that this pairing has already been performed, so that any asymptotic quantity extracted from the flavor-brane profile, like $\chi^{(-3)}$ in eq. \eqref{qqchi}, is understood to be replaced by twice its real part.\footnote{For a more in-depth analysis of the pole-pairing procedure and the related physics, see the discussion in \cite{Kaminski:2009dh}.}

\section{Induced expectation value and inverse magnetic catalysis}
\label{sec:IndExpResults}

\begin{figure}[htbp]
    \centering
    \includegraphics[width=0.75\textwidth]{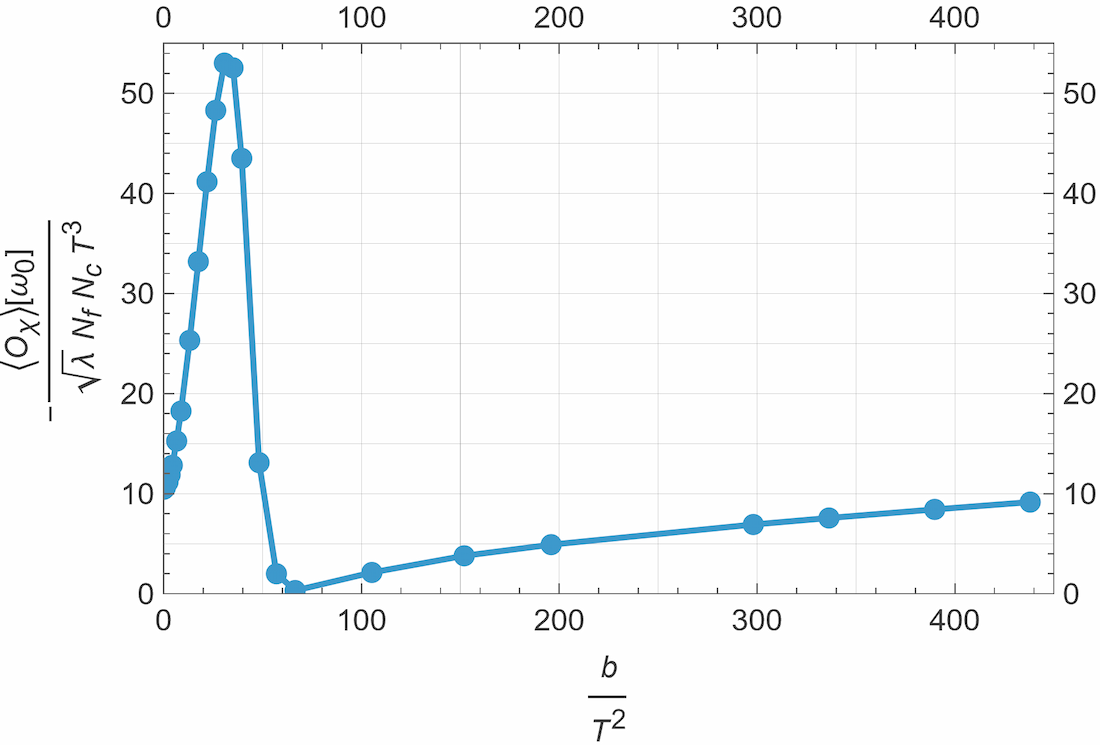}
    \caption{Full contribution to the induced expectation value of the scalar operator as a function of the magnetic field $b/T^2$, where both the effects of $b$ in the background geometry and the perturbation equation are considered.}
    \label{fullRes}
\end{figure}

We now present the results for the induced expectation value of the scalar operator. In figure \ref{fullRes}, where the dimensionless quantity $-8\left\langle \mathcal{O}_{\chi}\right\rangle/\sqrt{\lambda}N_f N_c T^3$ is presented as a function of $b/T^2$, we show that the full contribution initially increases, enhancing $\langle \mathcal{O}_{\chi}\rangle$, until it reaches a maximum at around $b/T^2=28\pi^2/9$. It then decreases over a range of magnetic field intensities, signaling a suppression of the expectation value, before reaching a minimum at approximately $b/T^2=20\pi^2/3$ and increasing once again as the magnetic field becomes stronger.

To isolate the valence contribution, we follow the prescription developed in section \ref{sec:ValenceSea}, where we proposed computing it by turning off the magnetic field in the background while retaining its explicit appearance in the equation governing the D7-brane perturbation. As discussed above, this causes the quasinormal modes to depend on $b$ and $k_z$ only through the kinetic momentum. To verify that the magnetic field had indeed been set to zero everywhere in the background as explicitly required, we numerically confirmed that the same results were obtained for any given value of ${k_K}^2/T^2$, independently of those particular to $b$ and $k_z$ used to construct it, with the magnetic-field dependence entering eq. \eqref{Seom} through the substitution ${\cal{E}}=\frac{1}{2}eb$.

\begin{figure}[htbp]
    \centering
    \includegraphics[width=0.75\textwidth]{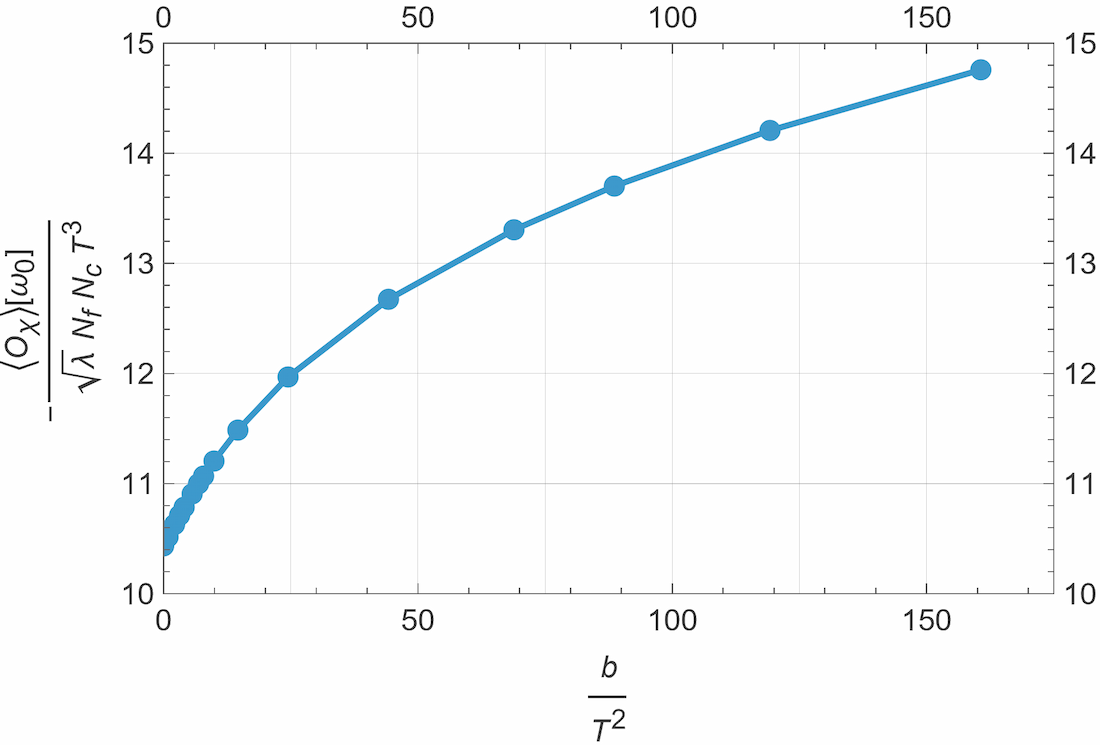}
    \caption{Valence contribution to the induced expectation value of the scalar operator as a function of the magnetic field $b/T^2$, where only the effect of $b$ on the perturbation equation through its explicit appearance is considered and the black D3-brane metric functions are used.}
    \label{ValRes}
\end{figure}

The resulting function, presented in figure \ref{ValRes}, increases monotonically throughout the explored range of magnetic field intensities and shows no indication of a change in this behavior. The valence contribution therefore acts exclusively to enhance $\langle \mathcal{O}_{\chi}\rangle$ over the considered range of $b$.

\begin{figure}[htbp]
    \centering
    \includegraphics[width=0.75\textwidth]{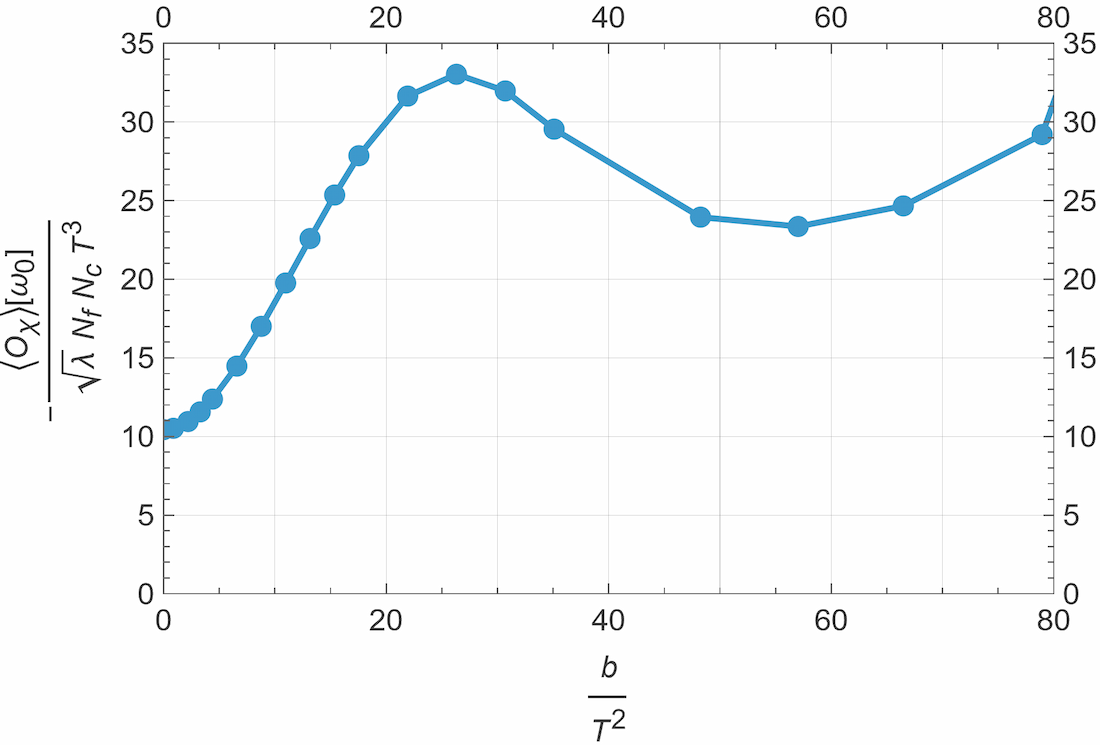}
    \caption{Sea contribution to the induced expectation value of the scalar operator as a function of the magnetic field $b/T^2$, where only the effect of $b$ on the background geometry is considered and is set to zero wherever it explicitly appears on the perturbation equation.}
    \label{SeaRes}
\end{figure}

In contrast, as shown in figure \ref{SeaRes}, the sea contribution exhibits the more intricate behavior observed in the full result, with an initial enhancement of $\langle \mathcal{O}_{\chi}\rangle$ up to approximately $b/T^2=8\pi^2/3$, followed by a suppression regime, before returning to an enhancing behavior at around $b/T^2=52\pi^2/9$. For reasons that will be clarified in appendix \ref{sec:AppA}, we were not able to compute this contribution for arbitrarily large values of the magnetic field intensity. Nevertheless, within the range over which we retained numerical control, we found no indication of an additional turning point.

\begin{figure}[htbp]
    \centering
    \includegraphics[width=0.75\textwidth]{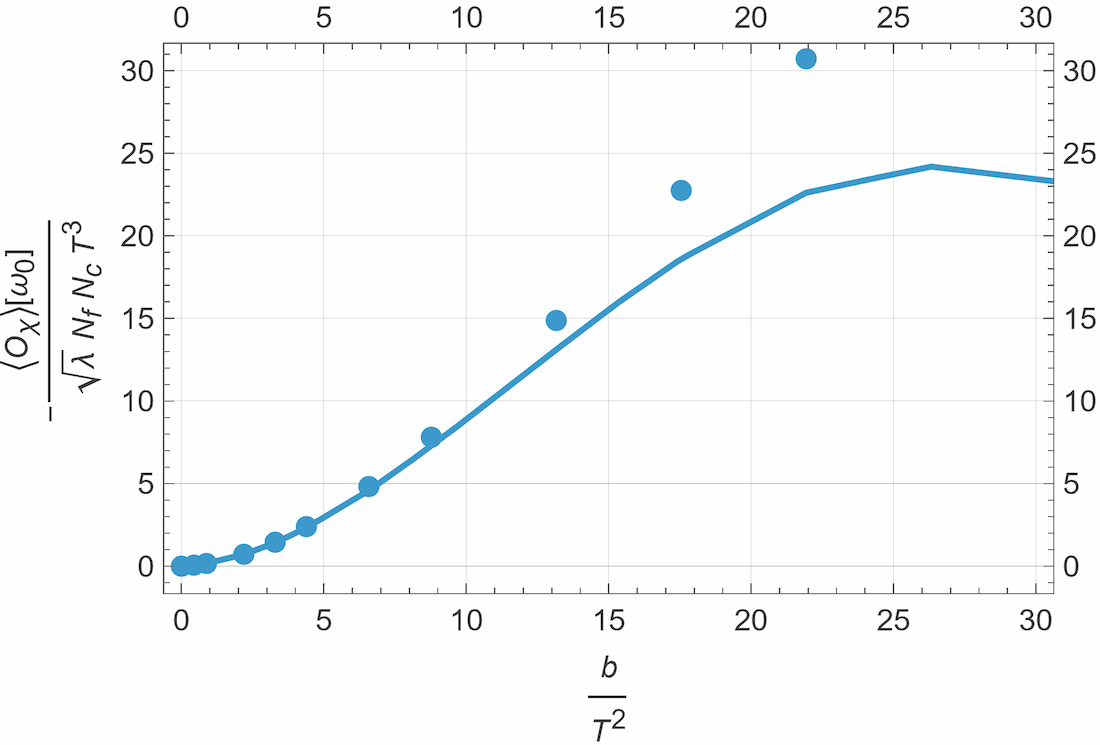}
    \caption{Comparison between the full contribution to the induced expectation value of the scalar operator and the sum of the sea and valence contributions, zoomed for smaller values of the magnetic field intensity.}
    \label{CompRes}
\end{figure}

Finally, we show in figure \ref{CompRes} that, for small values of the magnetic field intensity, the sum of the valence and sea contributions reproduces the full contribution to the induced expectation value. As the magnetic field increases, however, the two results begin to deviate from one another. This behavior when comparing the addition of the independent contributions with the full effect is anticipated in studies that introduce such separation \cite{Bruckmann:2013,DElia:2011koc}. In order to show this result and following the same procedure as in \cite{Bruckmann:2013}, the expectation value at $b=0$ was subtracted, eliminating additive divergences.

\section{Discussion and future work}

The main objective of the present work was to take advantage of a background with a fully backreacted magnetic field, and use the relevance that the presence of such field has in high energy collisions, to introduce a holographic prescription that distinguishes valence from sea effects in the QGP, and perform a first test of such prescription.

As part of the construction, we verified in section \ref{QPModel} that the modes used in our study follow the dispersion relation expected for a massless transverse field in a thermal theory, and that the inequality $|\operatorname{Im}[\omega]|\ll|\operatorname{Re}[\omega]|$ becomes increasingly well satisfied as the kinetic momentum grows, showing quasiparticle-like behavior. Opposite to the above, we showed that for these states the magnetic field has a destabilizing effect, suppressing the quasiparticle character of the modes. A similar suppression of quasiparticle character has been found to arise from interaction-induced spectral broadening in strongly coupled descriptions of the plasma \cite{Liu:2017qah,Liu:2016ysz}. If an analogous competition between the effects of momentum and the magnetic field was indeed present in the quark-gluon plasma that inspired our work, we would see that, on the one hand, the energy distribution placing a substantial fraction of the modes in the momentum range where they exhibit a quasiparticle-like behavior could help explain why quasiparticle-based effective descriptions can successfully be used to compute observables and describe experimental data in heavy-ion collisions \cite{Bratkovskaya:2011wp,Alqahtani:2017jwl}. On the other hand, the quasiparticle suppression due to the remarkably large magnetic field produced in heavy-ion collisions could contribute to the difficulties and challenges that appear when trying to isolate experimental signatures of quasiparticle constituents of the plasma \cite{DEramo:2012uzl,ALICE:2022wpn,ALICE:2024fip,Hulcher:2026dht}.

It is at this point that the importance of using our gravitational configuration becomes evident. It comes from our holographic prescription, which assumes that sea effects on the system are captured by the reaction of the background geometry to the magnetic field, along with its repercussions on the flavor brane embedding, while the valence contribution is encoded explicitly in the perturbation equations, provoking the necessity to create a scenario where such entities can be modified independently. That is exactly what a brane embedded in a background that includes a fully backreacted magnetic field provides, since it enables us to change the background to independently probe its effect on the perturbations of the embedding, or modify the value of the intensity of the field directly in the embedding equation to see what results of this action.

This procedure led to an equation describing the valence contribution to the excitations in which the spatial momentum and the magnetic field enter exclusively through their combination in the kinetic momentum, ${k_K}^2=eb+{k_z}^2$, for the lowest energy level, $n=0$. This particular dependence motivates our return to the study of spectral functions at zero spatial momentum, $k_z=0$, following our analysis of the damping rate and dispersion relations of the quasinormal modes.

With all of the above in hand, the numerical results for the induced expectation value of the scalar operator showed that, when both effects of the magnetic field—namely, its influence on the backreacted background and its explicit appearance in the perturbation equation—are simultaneously present, $\langle \mathcal{O}_{\chi}\rangle$ is initially enhanced, subsequently suppressed over an intermediate range of $b/T^2$, and finally enhanced once again beyond a certain value of the magnetic field intensity. This behavior qualitatively resembles that expected for the chiral condensate in LQCD in the context of magnetic catalysis and inverse magnetic catalysis \cite{Bruckmann:2013}. In our setup, however, the temperature is held fixed, so rather than identifying a critical temperature, the transition between these different behaviors is characterized in terms of the magnetic field intensity.

At the same time, we showed that the valence contribution enhances the induced expectation value throughout the studied range of magnetic field intensities, whereas the sea contribution reproduces the more intricate behavior of the full result, including a regime in which the expectation value is suppressed. This indicates that the direct effect of the magnetic field on the perturbation favors magnetic catalysis, while its effect through the backreacted background geometry is associated with both enhancement and suppression of the expectation value, and can therefore give rise to either magnetic catalysis or inverse magnetic catalysis. This qualitative behavior is consistent with the expectations from the LQCD literature \cite{Bruckmann:2013}.

The fact that the sea and valence contributions exhibit such distinct behaviors, while capturing different aspects of the full result, demonstrates the usefulness of implementing this distinction in the holographic setting, even if only static contributions associated with the sea sector were accessible within the particular configuration considered in this work. In addition, eq. \eqref{CVS} was presented in \cite{Bruckmann:2013} as an approximation, in the sense that the addition of both contributions is not expected to match the value of the condensate computed using all the elements except for at small values of $b$. Our holographic construction precisely corroborated such expectation.

It is also important to note that, while the quasiparticle analysis required the use of spectral functions with non-zero spatial momentum, the analysis for the induced scalar expectation value is done for vanishing spatial momentum $k_z=0$, making the results comparable with measurements that are typically performed in such conditions.

Even if the analysis above exhausts the scope of the current work, it also indicates at least two elements that require further investigation.

A topic that can be addressed using our construction is the effect of a magnetic field over the thermal mass \cite{Koothottil:2019}, in our case, of the quasiparticles dual to the modes we studied. This could be done performing a very precise calculation to extract the value of $m_T$ that better fits the asymptotic behavior in eq. \eqref{Asym} for different intensities of the magnetic field.

Another natural, though highly non-trivial, extension of the work presented here would involve two steps. The first is to consider the backreaction of the embedded flavor branes on the background, allowing for the direct computation of the full expectation value, including dynamical sea effects. The second is to consider embeddings with $m_q \neq 0$ beyond the perturbative regime, since the quark mass plays an important role in modulating which contribution, sea or valence, dominates \cite{Bruckmann:2013}.

\appendix
\section{Quasinormal-mode frequency analysis}
\label{sec:AppA}

Although this analysis does not constitute part of the main results, it is nevertheless useful to study how the magnetic field affects the quasinormal modes through the sea and valence contributions separately, and how these effects are reflected in the full result.

\subsection{Imaginary part of the frequency}

\begin{figure}[htbp]
    \centering
    \includegraphics[width=0.75\textwidth]{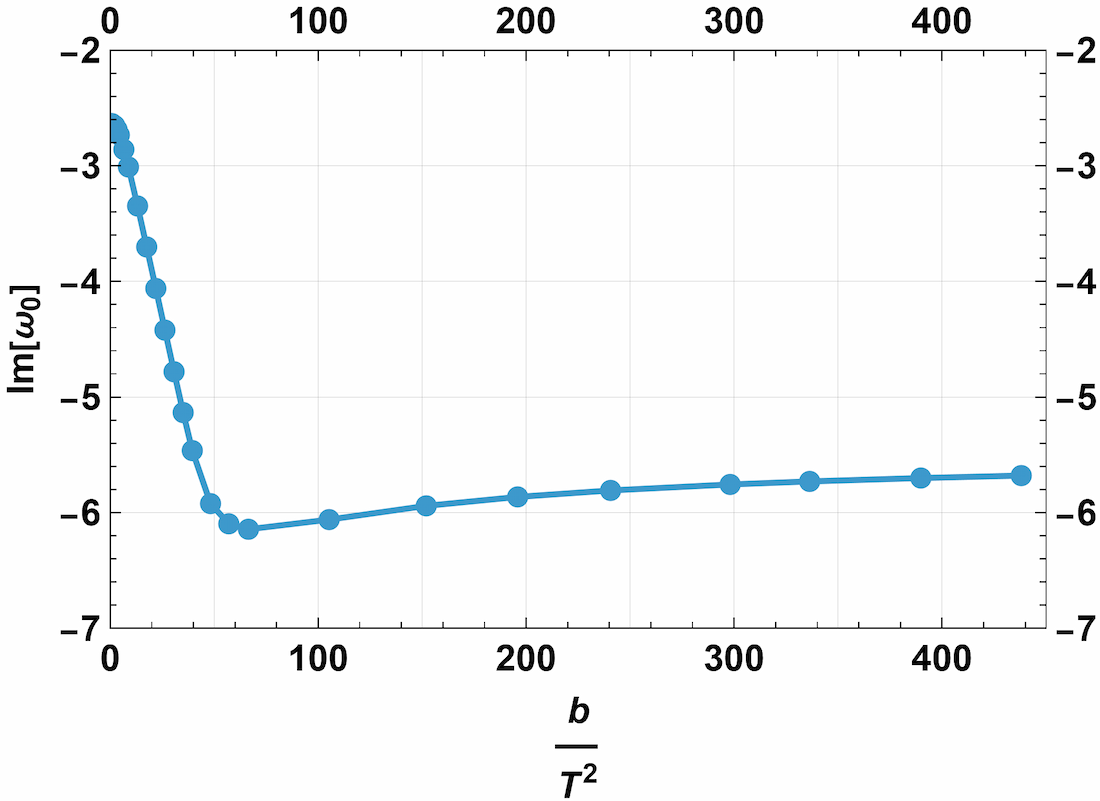}
    \caption{Imaginary part of the quasinormal-mode frequency as a function of the magnetic field $b/T^2$, including the effect of $b$ both on the background geometry and the perturbation equation.}
    \label{FullImFreq}
\end{figure}

With the aim above, figure \ref{FullImFreq} shows the imaginary part of the quasinormal frequency when the full effect of $b$ is included. At lower magnetic field intensities, the mode becomes increasingly damped as $b$ grows, until a turning point is reached, after which the damping gradually decreases.

\begin{figure}[htbp]
    \centering
    \includegraphics[width=0.75\textwidth]{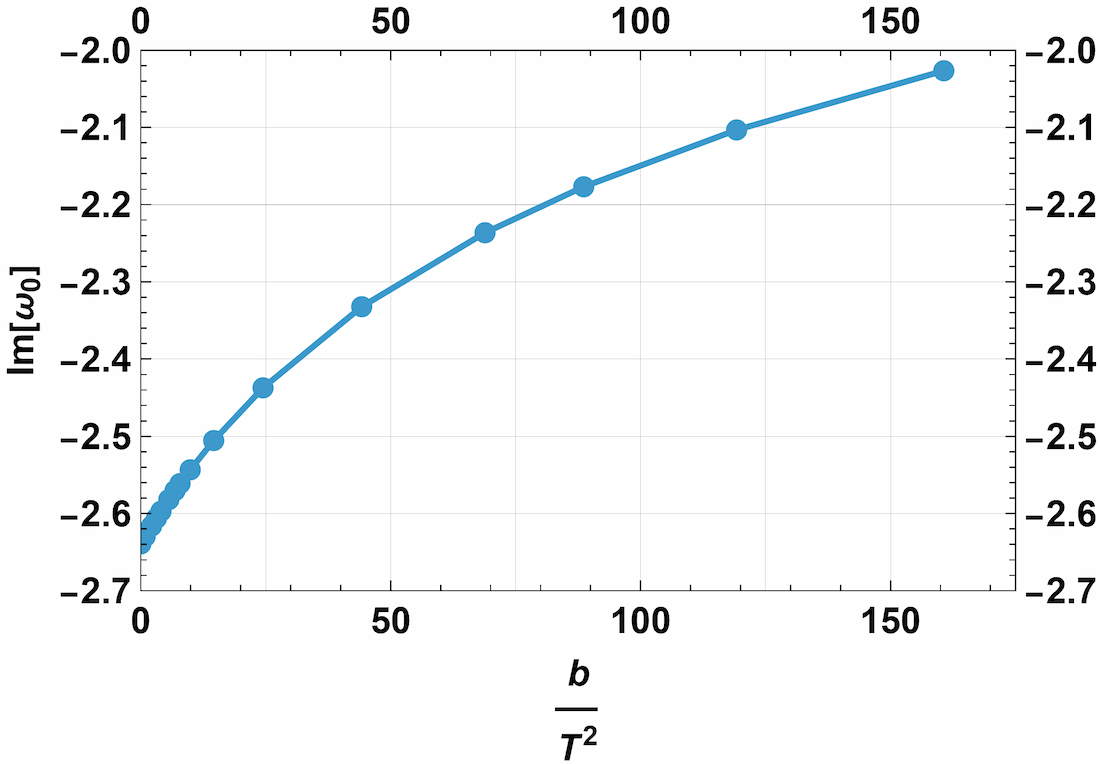}
    \caption{Imaginary part of the quasinormal-mode frequency as a function of the magnetic field $b/T^2$, including only the explicit effect of $b$ in the perturbation equation.}
    \label{ValImFreq}
\end{figure}

\begin{figure}[htbp]
    \centering
    \includegraphics[width=0.75\textwidth]{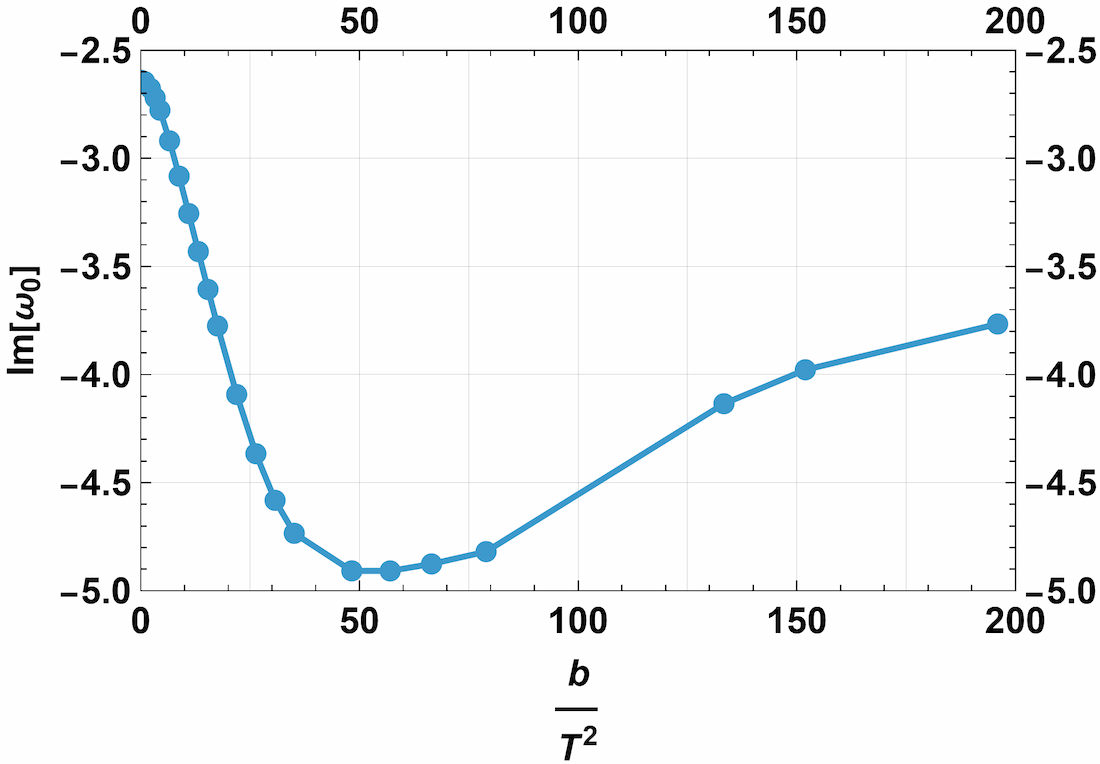}
    \caption{Imaginary part of the quasinormal-mode frequency as a function of the magnetic field $b/T^2$, including only the effect of $b$ through the backreacted background geometry.}
    \label{SeaImFreq}
\end{figure}

In figure \ref{ValImFreq}, the magnitude of the imaginary part of the quasinormal frequency for the valence contribution is shown to decrease monotonically, indicating that the direct effect of the magnetic field on the perturbation is to reduce the damping of the quasinormal modes. Meanwhile, figure \ref{SeaImFreq} shows that, for the sea contribution, the magnitude of the imaginary part initially increases and subsequently decreases. The effect of the magnetic field through the backreacted background geometry is therefore responsible for the regime of enhanced damping observed in the full result.

\subsection{Real part of the frequency}

\begin{figure}[htbp]
    \centering
    \includegraphics[width=0.75\textwidth]{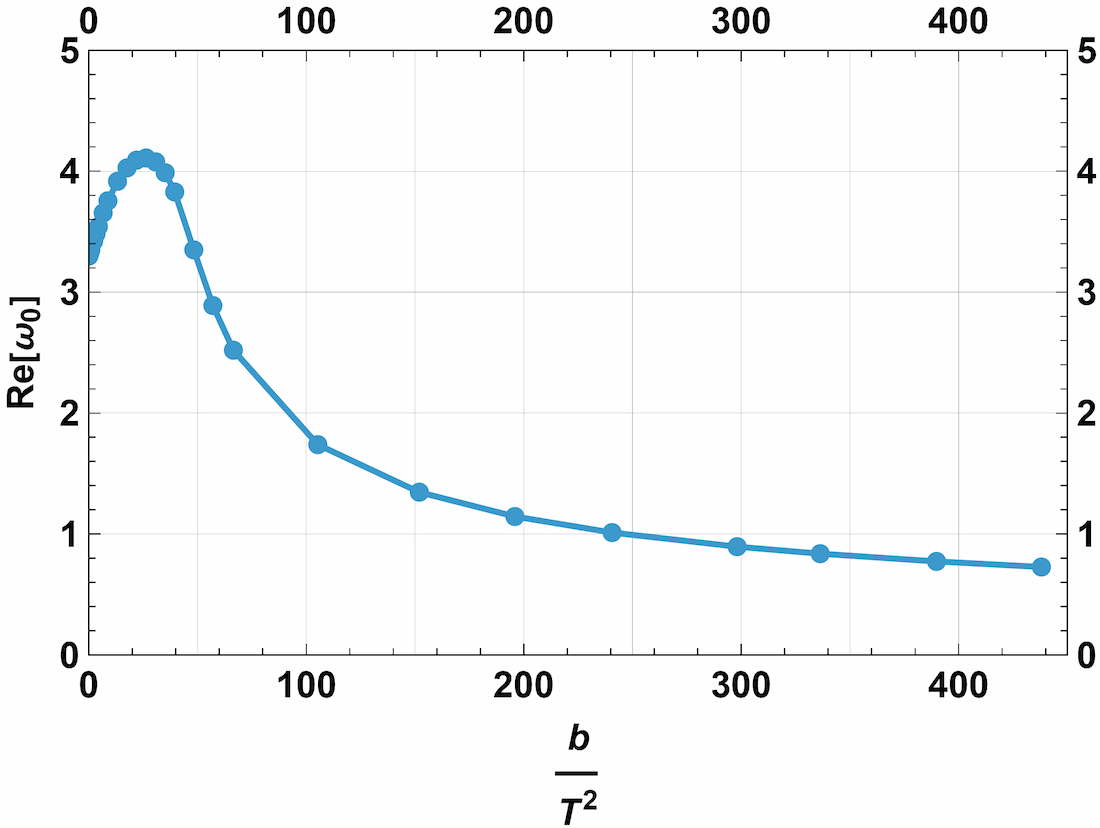}
    \caption{Real part of the quasinormal-mode frequency as a function of the magnetic field $b/T^2$, including the effect of $b$ both on the background geometry and the perturbation equation.}
    \label{FullReFreq}
\end{figure}

\begin{figure}[htbp]
    \centering
    \includegraphics[width=0.75\textwidth]{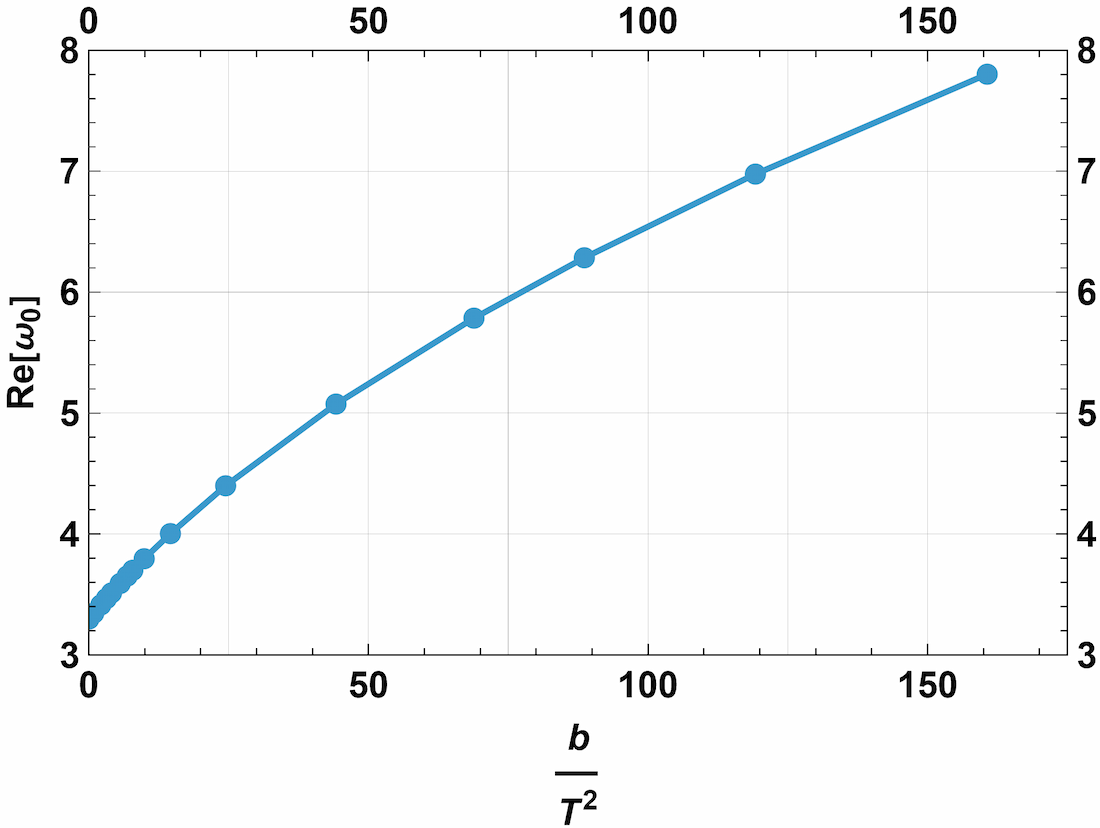}
    \caption{Real part of the quasinormal-mode frequency as a function of the magnetic field $b/T^2$, including only the explicit effect of $b$ in the perturbation equation.}
    \label{ValReFreq}
\end{figure}

Figure \ref{FullReFreq} shows that, when the full effect of the magnetic field is included, the real part of the quasinormal frequency initially increases with the field intensity and subsequently decreases. In contrast, the valence contribution shown in figure \ref{ValReFreq} increases this quantity monotonically, indicating once again that the non-monotonic behavior of the full result originates from the effect of $b$ through the backreacted background geometry. The fact that the valence contribution simultaneously increases the real part of the frequency and suppresses the damping rate indicates that it tends to drive the modes toward the quasiparticle regime.

\begin{figure}[htbp]
    \centering
    \includegraphics[width=0.75\textwidth]{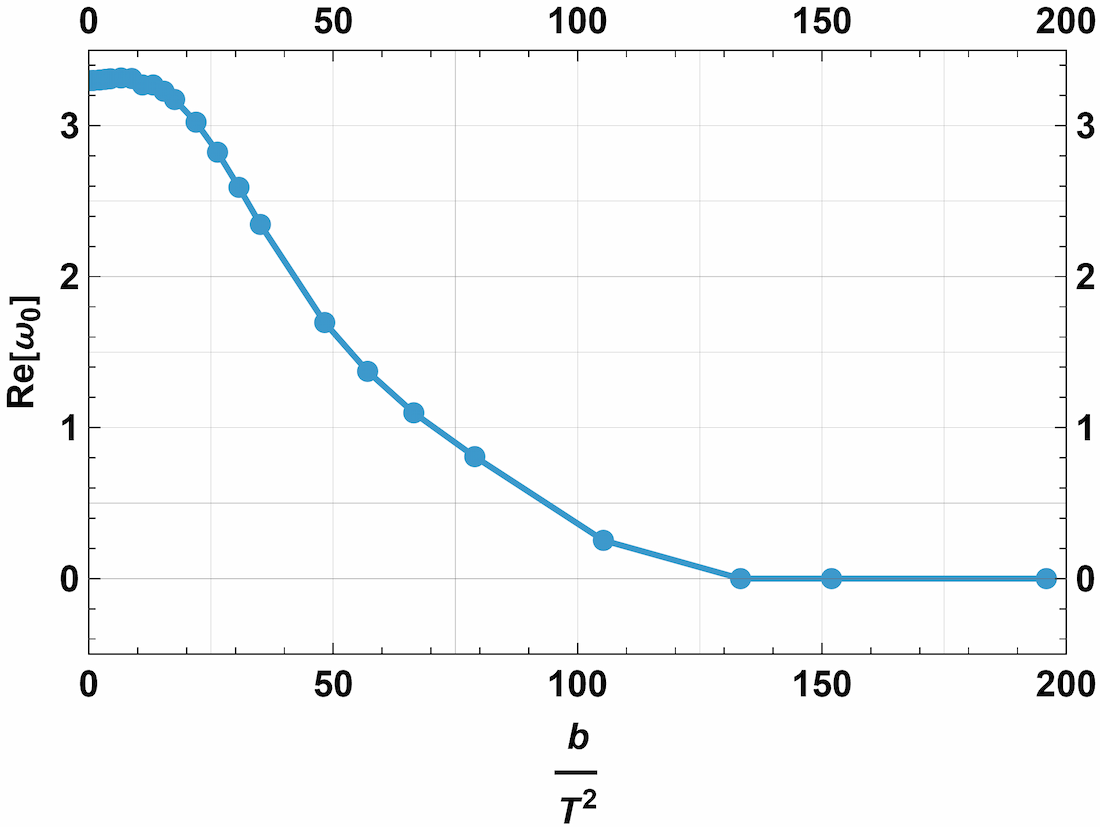}
    \caption{Real part of the quasinormal-mode frequency as a function of the magnetic field $b/T^2$, including only the effect of $b$ through the backreacted background geometry.}
    \label{SeaReFreq}
\end{figure}

Figure \ref{SeaReFreq}, which shows the real part of the quasinormal frequency for the sea contribution, corroborates the previous observation that the turning point and subsequent decrease observed in the full result originate from this contribution. In particular, as the magnetic field intensity increases, the real part of the frequency tends toward zero, indicating that the corresponding quasinormal modes become increasingly dominated by their imaginary component. The sea contribution therefore tends to drive the modes away from the quasiparticle regime and toward an overdamped, non-propagating one, where hydrodynamic poles could be found \cite{Kaminski:2009dh}.

It is this behavior that limits our study of the sea contribution, since, as the modes approach the purely imaginary regime, we appear to lose numerical control over the asymptotic expansion coefficients of the embedding function.

\clearpage
\acknowledgments

We are very thankful to Alejandro Ayala for meaningful discussions about the thermal dispersion relations. The work of D.K. is supported by CONAHCyT MSc. grant. All the plots in this paper were generated using Wolfram Mathematica.

\bibliographystyle{JHEP}
\bibliography{AdSbib}

\end{document}